# Improve the Practice of Software Development in India by Having a Software Development Career Track in Indian CS & IT Academia

Ravi S. Iyer
Software Consultant, Puttaparthi, Andhra Pradesh, India
ravi@raviiyer.org

December 21st, 2012

**Abstract**

*Many, but not all, Indian CS & IT academics tend to have a focus on theory and research. They do not give much importance to the practice of software development. This paper proposes an additional software development career track for Indian CS & IT academics different from the existing research oriented career track. A measure of software contribution record is suggested. It opines that adoption of such changes to academic regulations will result in significant improvement of software development skill set in Indian CS & IT academia which, in turn, will result in better software development skill set in Indian CS & IT graduates.*

*Note: The review remarks for this article by a noted international academic publication focused on CS education and the response of the author are provided in Appendix A.*

# Introduction

> *The author is a Physics Graduate (and Physics Masters drop-out) from India who was industry-trained and later self-taught in software development. He worked in the international software industry (US, Europe, Japan, South Korea, India etc.) developing systems as well as applications software (CS & IT) for over 18 years after which he retired from commercial work. He later, mainly as an "honorary faculty/visiting faculty", offered free service of teaching programming courses (lab. courses) and being a "technical consultant" for student projects in a Maths & Computer Science department of a deemed university in India for 9 years. This paper is mainly based on this experience of the author.*

The rather odd reality of the vast number of CS & IT departments of universities & colleges in India is that the majority of the teachers in these departments focus on theory and research publications but do not give much importance to practical areas like Software Design and Programming or Coding. Therefore the practice of software development is quite poor in most Indian CS & IT departments. The sections below give references to support these statements.

# Peer Reviewed Academic Literature Sources on Poor Software Development Skill Set in Indian CS & IT Academia

Mahanti et al., 2005, state that in India, "Software engineering does not yet have an independent curriculum with enough durable, codified content to justify a separate undergraduate curriculum." [20]. They further state that in India, "Limited exposure to industry problems, inability to adapt course curricula to dynamic industry requirements, limited exposure to latest tools & techniques, inability to enter into emerging areas, rigid and outdated course curricula, (repetition omitted), poor industry linkages, little real-life case studies, little scope for creative learning are some of the drawbacks in the software education system in the universities."

Garg et al., 2008, conducted a survey of major software services companies in India and reported that the training programs of these companies include retraining on programming and Software Engineering (SE) as Indian academia is not able to impart these skills to the level that they expect [21]. They further state that they studied the publicly available syllabi related to SE for a large number of Indian universities and found that the focus is on theoretical aspects and "Practical aspects, best practices, recent developments are not included and students rarely get a chance for application of the knowledge and skills they learned."

# General Public Views on Poor Software Development Skill Set in Indian CS & IT Academia

The author believes that Indian CS & IT academics should also listen to views of students and others like the news media and teacher blogs on this matter. While these views may be contested as not having been validated by means of an academic/scientific peer review process the author is of the opinion that the almost unanimous voice of the general Indian public must be given some importance. The Indian public naturally expects that Indian CS & IT academics will prepare Indian CS & IT students to contribute mainly as competent software developers to the explosively growing software systems that pervade many aspects of modern life in India and the world.

Most of the students graduating out of Indian CS & IT academia end up having somewhat strong theoretical and, at times, research-oriented skills but being poor in Software Design and Programming [1, 12, 13]. The problem lies not with the CS & IT academics but with the Indian CS & IT academic system which provides career growth mainly for research output and largely ignores software contribution output [1].

A study of over 50,000 engineers who graduated in 2011 in India, very alarmingly states, "The percentage of ready-to-deploy engineers for IT jobs is dismally low at 2.68%" [19]. It further, alarmingly again, states, "An economy with a large percent of unemployable qualified candidates is not only inefficient, but socially dangerous." A Bangalore, India industry organization is planning to set up a task force to have an interface between IT industry and academia to restructure academic courses to ensure that IT graduates have skills desired by industry [18].

For this rather odd situation, where most Indian CS & IT academics/teachers do not have a strong software development skill set, to improve, Indian academic regulations should provide career advancement incentive for CS & IT academics who have a software contribution record [4]. Further, students should be informed of the software contribution record of faculty of CS & IT departments by making it mandatory for CS & IT departments to put up such information on its web site [5].

Teaching excellence in Indian CS & IT academia does not seem to be given much importance. It may be due to an obsession with research as there seems to be no significant rewards or recognition for teaching excellence whereas research excellence gives recognition, even fame at times, and career advancement [6].

How can we improve software development teaching standards in Indian CS & IT academia? The sections below examine Indian academic regulations with this objective.

## A Short Introduction to The Indian CS & IT Academic System

The University Grants Commission (UGC) is the apex academic body of India [10]. "The UGC has the unique distinction of being the only grant-giving agency in the country which has been vested with two responsibilities: that of providing funds and that of coordination, determination and maintenance of standards in institutions of higher education." [10a]. The All India Council of Technical Education (AICTE) [10d] is a professional council which operates under UGC umbrella. The UGC website states about AICTE, "The council is authorized to take all steps that are considered appropriate for ensuring coordinated and integrated development of technical education and for maintenance of standards." [10b].

"The UGC serves as a vital link between the Union and State Governments and the institutions of higher learning." [10c]. The UGC regulations/norms for appointment of academics [7] has a significant influence on the career of Indian academics of all academic streams including CS & IT who are employed in any higher education institution regulated by UGC or its professional councils like AICTE. But the extent of influence may vary depending on whether the educational institution receives government aid (funds) or not. The entry level position for a regular teacher post (as against a Teaching Assistant post) is the Assistant Professor position. The other teacher positions are Associate Professor and Professor.

In India the degrees awarded for software education by UGC/AICTE regulated academia are mainly termed as Computer Science (CS) or Information Technology (IT) degrees with an additional variant of Computer Applications degrees [15]. The Software Engineering (SE) degree/program is not well known in Indian academia. Most universities offer Software Engineering as a course along with other courses in their CS (and IT) curriculum [21].

The elite Indian Institute of Technology (IIT) institutions are independent of UGC & AICTE but they cater to only a small percentage of technical students in India. There are also thriving private software education/training institutes with a nation-wide presence sustained over decades but their certifications are different from the CS & IT degrees offered by UGC/AICTE recognized universities. This paper limits itself to UGC/AICTE recognized CS & IT educational institutions.

## UGC Appointment & Promotion Regulations for Music & Dance Discipline

Music, including the vocal art of singing, & Dance are performing arts. The teacher of these arts must be a capable performer first and should also have adequate theoretical knowledge.

This aspect of Music & Dance being a practice-oriented discipline is reflected in UGC [10] regulations for appointment of Assistant Professor, Associate Professor & Professor for Music & Dance discipline on Pages 7 - 9 of its regulations for appointment of teachers [7]. The author presumes that the regulations for promotion for Music & Dance discipline teachers will be on similar lines. These UGC regulations for Music & Dance discipline can be summarized as follows:

For the Assistant Professor post, the candidate should conform to standards similar to regular disciplines like Physics and Mathematics which are: Master's degree with 55 % Marks + NET/SLET/SET (National Eligibility Test, State Eligibility Test etc.) qualification; PhD, adequate research publication record etc. come into play for higher posts of Associate Professor and Professor.

    OR

For the Assistant Professor post, the candidate should have studied under noted traditional masters, be a high grade artist of AIR/TV (Radio/Television) and have adequate theoretical knowledge; years of performance, participation in national/international seminars/workshops etc. come into play for higher posts of Associate Professor and Professor.

Specifically, practical expertise of the performer is recognized and formal academic degree qualification in Music or Dance as well as research publications are not necessary.

## Suggestion of Two Tracks for CS & IT Academics: Research Oriented and Software Development Oriented

The software development discipline is a very practice oriented discipline. Design & programming (coding) are vital skills. Of course, theoretical background is important but theory not backed by competent design & programming skill will make a software development practitioner/professional as incompetent as a musician or dancer who knows theory well but is not competent in performing music or dance.

Research is also vital for the software field. It is research that creates fundamental advances in the Computer Science & Information Technology (CS & IT) fields. Without research, the great and revolutionary force of the Internet would not have been created. This single example, itself,

Page **4** of **21**

of the benefit of research shows how critical it is for progress in the software field. There are many, many other areas of CS & IT research that are of great importance to the software field.

In the author's opinion, we need both types of CS & IT teachers - practice oriented software development teachers & research oriented teachers. A very few teachers may excel at both, software development as well as research. But that will, in all probability, be a numerically insignificant minority among the huge number of CS & IT academics in the country.

As of now, UGC appointment & promotion regulations do not differentiate between CS & IT disciplines and disciplines like Physics & Mathematics [7]. The author could not find an equivalent regulations document for AICTE on its website but the general impression is that AICTE follows regulations similar to UGC in this regard. There is no incentive for practice-oriented software development teachers resulting in the majority of CS & IT academics being theory and research-oriented with not-so-strong software development skill set/knowledge. When the software development teacher himself is not so knowledgeable about software development the probability of students being taught software development skills well is very low.

The author suggests that UGC & AICTE regulations for appointment and promotion of CS & IT academics be modeled on the lines of that of Music & Dance discipline. The author would not like to get into the debate of CS as science vs. CS as art. His emphasis is on the software development part of CS & IT being a very practice oriented discipline like Music and the performing arts. Further, in a tightly regulated system like Indian academia, precedent for any suggested change makes it easier to consider the change. Since the regulations for Music and the performing arts already have a mechanism to cater to both the research oriented teachers and the practice oriented teachers, the author considers it appropriate to refer to the precedent and suggest a similar mechanism for Indian CS & IT teachers.

There should be two tracks for CS & IT academics - the current one for research oriented academics and another for practice oriented software development academics. Like the measure for competence in research for the (research oriented) CS / IT academic is the research publication record, the measure for competence of the software development CS / IT academic should be the quality and quantity of her open source software contribution record.

# Suggested Changes to Teacher Eligibility Tests (NET/SLET/SET) for CS & IT Disciplines

According to UGC regulations [7] the minimum requirements for an Assistant Professor appointment in engineering and technology discipline (which includes CS & IT) are a first class Master's degree in the appropriate branch of engineering and technology and qualifying in the teacher eligibility tests (NET/SLET/SET) [11]. AICTE seems to have watered down the requirement of Master's degree to a Bachelor's degree in engineering/technology discipline probably due to paucity of adequately qualified candidates applying for the Assistant Professor position. The teacher eligibility test is waived for candidates who have been awarded a Ph.D. degree [7].

Page **5** of **21**of the benefit of research shows how critical it is for progress in the software field. There are many, many other areas of CS & IT research that are of great importance to the software field.

In the author's opinion, we need both types of CS & IT teachers - practice oriented software development teachers & research oriented teachers. A very few teachers may excel at both, software development as well as research. But that will, in all probability, be a numerically insignificant minority among the huge number of CS & IT academics in the country.

As of now, UGC appointment & promotion regulations do not differentiate between CS & IT disciplines and disciplines like Physics & Mathematics [7]. The author could not find an equivalent regulations document for AICTE on its website but the general impression is that AICTE follows regulations similar to UGC in this regard. There is no incentive for practice-oriented software development teachers resulting in the majority of CS & IT academics being theory and research-oriented with not-so-strong software development skill set/knowledge. When the software development teacher himself is not so knowledgeable about software development the probability of students being taught software development skills well is very low.

The author suggests that UGC & AICTE regulations for appointment and promotion of CS & IT academics be modeled on the lines of that of Music & Dance discipline. The author would not like to get into the debate of CS as science vs. CS as art. His emphasis is on the software development part of CS & IT being a very practice oriented discipline like Music and the performing arts. Further, in a tightly regulated system like Indian academia, precedent for any suggested change makes it easier to consider the change. Since the regulations for Music and the performing arts already have a mechanism to cater to both the research oriented teachers and the practice oriented teachers, the author considers it appropriate to refer to the precedent and suggest a similar mechanism for Indian CS & IT teachers.

There should be two tracks for CS & IT academics - the current one for research oriented academics and another for practice oriented software development academics. Like the measure for competence in research for the (research oriented) CS / IT academic is the research publication record, the measure for competence of the software development CS / IT academic should be the quality and quantity of her open source software contribution record.

# Suggested Changes to Teacher Eligibility Tests (NET/SLET/SET) for CS & IT Disciplines

According to UGC regulations [7] the minimum requirements for an Assistant Professor appointment in engineering and technology discipline (which includes CS & IT) are a first class Master's degree in the appropriate branch of engineering and technology and qualifying in the teacher eligibility tests (NET/SLET/SET) [11]. AICTE seems to have watered down the requirement of Master's degree to a Bachelor's degree in engineering/technology discipline probably due to paucity of adequately qualified candidates applying for the Assistant Professor position. The teacher eligibility test is waived for candidates who have been awarded a Ph.D. degree [7].

The present teacher eligibility test (NET/SLET/SET) for CS / IT academics is a paper only test (though the syllabus includes C/C++ & SQL [11]) due to which an aspirant can become eligible to be appointed as Assistant Professor without having good practical software development skills! That may be acceptable for a research track CS / IT teacher. But it is unacceptable for a software development track CS / IT teacher. Aspirants who do not have good practical software development skills should NOT be appointed as software development track CS / IT Assistant Professors (or other grade Professors).

A new teacher eligibility test for software development track CS / IT academics should be introduced which will have a 50% weight-age practical test (on computer) involving programming and some amount of design, and 50% weight-age on theory. This will ensure that software development track teacher-aspirants will have to be reasonably good in both theory and practice aspects of CS / IT.

# Allow Movement from Research Track to Software Development Track & Vice-Versa

A CS / IT academic should be able to switch track from research oriented to software development oriented if her software contribution record is appropriate. Similarly a software development oriented CS / IT academic should be able to switch track to research oriented if his research publication record is appropriate. Some CS / IT academics may have a respectable research publication record as well as a respectable software contribution record which would be a wonderfully balanced contribution record.

# How Do We Measure a Software Contribution?

This will have to be evolved over time. Software industry bodies in India like NASSCOM & CSI [8] (other countries would have other such bodies) can arrive at norms for evaluating an academic software contribution which can be updated at appropriate intervals to reflect the rapidly changing software practice. The author suggests the following for measuring (and sharing) the academic software contribution:

1. It should be open source allowing any person to download the software and use it, examine it or modify it.
2. Industry professionals should "peer review" the candidate academic software contribution using norms provided by industry bodies like NASSCOM or CSI and decide whether it is of requisite quality & quantity to be considered as a "peer reviewed" academic software contribution. Note that the contribution can be a single author contribution or a multiple author contribution like academic publications can be single author or multiple author.
3. Over time, an impact factor similar to one used by scientific journals [9] can be evolved for a "peer reviewed" software contribution. Extent of usage of software can be considered for this impact factor like citations are considered in arriving at a scientific journal's impact factor.

4. To make it difficult for contributor-aspirants to fake, plagiarize or wrongly influence peer review of software contributions, any "peer reviewed" software contribution should be open to challenge by suitable industry professionals or academics. As the software will be downloadable including its source, a challenger will be in a position to study the contribution in depth and challenge its acceptance as a "peer reviewed" contribution. The challenge can be decided by an industry body like NASSCOM or CSI appointed referee.
5. All these "peer reviewed" open source academic software contributions should be properly listed and organized in a web based repository which is openly accessible.

Involving the software industry in this "peer review" of academic software contributions may go a long way in reducing the huge academia-industry disconnect in the software field today.

# Concern of Research Rigour Being Watered Down

One concern may be that the research rigour of CS & IT departments will get diluted by having practice oriented software development track teachers. Well, we need a balance. CS & IT departments should have the "right" balance of research oriented teachers & software development oriented teachers. The "right" balance for a research-intensive department could be 80 % research oriented teachers and 20 % software development oriented teachers. In contrast, the "right" balance for a teaching-intensive department could be more like 50 % software development oriented teachers and 50 % research oriented teachers. Please note that the software development oriented teacher has to be good at theory too and has to prove his theoretical knowledge by clearing the software development track NET/SLET/SET exam.

# Allow Industry-Trained & Self-Taught Professionals to Become CS & IT Teachers By Clearing Teacher Eligibility Tests (NET/SLET/SET)

The software industry has a huge number of industry-trained and self-taught professionals who do not have a CS / IT academic qualification. Some of the biggest icons of the software industry who are world-famous like Bill Gates (Microsoft), Late Steve Jobs (Apple) and Mark Zuckerberg (Facebook) are/were self-taught. India, in particular, has a vast number of industry-trained software professionals who come from various disciplines in engineering, science, management, commerce & even arts. UGC & AICTE must recognize this reality of the software/CS & IT fields and allow interested industry-trained and self-taught professionals with significant number of years of experience in the software industry to become regular (paid) CS & IT software development academics. For such software industry professionals the requirement of a Master's degree in CS / IT should be waived like it is waived for the performing artist track in Music & Dance discipline. But the self-taught software professional MUST prove his/her capability by passing the software development track eligibility test (NET/SLET/SET) which will test both his/her theoretical knowledge as well as practical competence.

Please note that UGC regulations (and AICTE regulations too, it is presumed) allow for an "outstanding professional" of a field to be appointed as a Professor. The above mentioned suggestion is for those who are not eminent but are knowledgeable & competent industry-trained and/or self-taught software professionals.

## Industry Professionals as Visiting Faculty/Industry Consultants

Industry professionals who are not NET/SLET/SET qualified nor possess a PhD but are offering free/honorary teaching service may be accommodated as visiting faculty/industry consultants if their knowledge and skill-set are found competent by university/college & department administrative authorities. Such industry professionals who offer their services to a university/college regularly may be an insignificant minority of the CS & IT teachers of the country. They may be treated as exception cases.

## A Brief Look at Software Engineering Education, Certification and Professional Licensure in USA and Some Other Countries

The author has direct exposure to only Indian software education academia and so has focused on it for most of this article. However, it was felt that mention of software education practices in some other countries would give a larger, international perspective. So he did a small literature survey to study efforts made to ensure good software development practice in software education in USA and some other countries, and extended it to cover certification and professional licensure. The study focuses more on Software Engineering (SE) degree programs than Computer Science (CS) degree programs. Judging what aspects of this small study report could be useful in Indian environment may ideally need somebody who has direct exposure to software education field in both India and other countries like USA. This author leaves those aspects for others to consider, if they find it worthy of consideration.

SE is an established program in USA academia distinct from a CS program [22]. The SE2004 volume gives guidelines for a SE curriculum and its website indicates that it was an exhaustive effort at improving SE education quality in the USA, UK, Australia, Canada, etc. [23]. Accreditations of SE (and other engineering & technology) programs are conducted by organizations specializing in accrediting technical education. Lethbridge et al., 2007, give details of SE programs and their accreditation in USA, Canada and UK [22].

At a USA institute, "software engineering is a five year program, with students graduating with the equivalent of almost a full year of work experience." and there is collaboration between various companies and the institute on projects as part of the SE education program [25]. Its website states that its senior projects involve a team of 4 to 5 students working on challenging, real-world software issues for companies & organizations and results in a functional software tool ready to be used by the organization [26].

Stroustrup and others have adopted a "software curriculum" in a CS program with an aim to produce 'software professionals (for some definition of "professional")' and reported largely positive results from it [3].

The IEEE Computer Society offers certification of graduates as well as self-taught software development professionals by conducting certification exams [24]. It's "Certified Software Development Associate (CSDA)" certification/credential "is intended for graduating software engineers and entry-level software professionals and serves to bridge the gap between your educational experience and real-world work requirements" [24a]. Its more advanced "Certified Software Development Professional (CSDP)" certification/credential "is intended for mid-career software development professionals that want to confirm their proficiency of standard software development practices and advance in their careers" [24b]. IEEE Computer Society claims that its certification programs are "industry standard measurements of fundamental software engineering practices" and so are different from vendor-specific & product-specific certifications [24c].

Land et al., 2012, argue that current circumstances are favorable for formal certification in software engineering to be considered and state that there is growing support for IEEE CSDA and CSDP in both industry and academia [27]. They further state that these certifications are based on the IEEE Computer Society's Guide to the Software Engineering Body of Knowledge (SWEBOK) [28] which is followed by the CS and SE programs of some colleges and universities.

Laplante, 2012, mentions that 10 states of USA may soon be requiring licensure for software engineers working on systems related to "public health, safety, and welfare" [27a]. He further mentions that the professional licensure requirements for software engineers will be similar to those of other engineering professions in the states of the USA and that most components of such licensure requirements/exams for software engineers are already in place with a final component expected to be available in April 2013. However, Miller, 2012, suggests that the enthusiasm for professional licensure of software engineers be tempered with caution [27b]. He states, "Questions about professionalism and licensing in IT have a complex, international history."

Mead, 2009, gives a timeline of SE education in USA and some other countries [29], notably:
- 1980's seeing the first conference on SE education;
- 1990's seeing first class graduating with Master of Software Engineering (MSE) degree of Carnegie Mellon University, undergraduate SE programs in other universities and its accreditation, growth of industry-university collaborations, a joint committee of ACM & IEEE Computer Society being formed to promote SE as a profession, licensing being introduced by US state of Texas and a lot of controversy being generated over licensing that "continues to this day", distance learning enabling global SE education;
- 2000's seeing IEEE Computer Society adopting SWEBOK and offering CSDP certification, many universities offering international SE programs and SE education track being introduced in other conferences besides CSEET.

# Critical Views on Software Education in USA and Some Other Countries

The author felt it appropriate to share some critical views on software education in USA and some other countries from academic and general public sources.

Stroustrup, 2010, has argued that "fundamental changes to computer science education are required to better address the needs of industry", and shows the disconnect between CS academia and industry [2]. Parnas, a veteran SE academic, in an ACM Fellow profile interview in 1999, states, "Most students who are studying computer science really want to study software engineering but they don't have that choice. There are very few programs that are designed as engineering programs but specialize in software." [30]. He also states that the term software engineering is often confused with project management techniques.

Mark Tarver, who taught in UK CS academia prior to 2000, is harshly critical of programming skills of UK final year project CS graduate students who confessed to not being able to do any programming. He is also harshly critical of UK CS education in general [14].

A student, 2010, captured the feelings of the student community when he wrote, "I'm graduating with a Computer Science degree but I don't feel like I know how to program" and tried to seek advice from a professional programmer forum [16]. A USA employer/interviewer, 2011, who has hired dozens of C/C++ programmers, stated, "A surprisingly large fraction of applicants, even those with masters' degrees and PhDs in computer science, fail during interviews when asked to carry out basic programming tasks" [17].

# Conclusion

If the practice oriented software development career track, as suggested in this paper, is introduced in UGC & AICTE regulations for appointment and promotion of Indian CS & IT academics then, over time, we will have a healthy mix of both research oriented as well as software development oriented Indian CS & IT academics. We may even have significant number of software development experts from the software industry moving to Indian CS & IT academia. What a boon that will be for boosting the software development skill set of Indian CS & IT academia! It will also dramatically reduce the huge academia-industry gap that plagues the Indian software field today.

These changes, in turn, will, at least for the teaching-intensive Indian CS & IT departments, result in graduates & post-graduates of CS / IT having a good balance of theory and practice of software development with some appreciation for the research angle of CS / IT as well. Some of these graduates/post-graduates may choose to pursue research by doing a PhD in CS / IT. Some may become CS / IT academics who will be more knowledgeable about practical software development than is the case now. The majority of them will typically take up industry software development jobs for which they will be far better equipped with the required software development practice skill set than they are now.

# Acknowledgements

The author's software industry and CS doctoral student friends have provided valuable contributions to the author's Indian CS & IT academic reform activism blog: http://eklavyasai.blogspot.in/p/table-of-contents.html. These interactions greatly encouraged the author to attempt this rather daunting task of making a case for a software development career track in Indian CS & IT academia to improve the practice of software development in India. However, the author would like to clarify that the views in this paper are his individual views. The author thanks the reviewers of a noted academic publication focused on CS education for their critical comments which led the author to limit most of his views to the Indian context as that is what the author has studied and experienced, strengthen the article with more peer reviewed academic references and broaden its view with a brief study of software education in USA and some other countries.

9. Impact Factor of Scientific journals: A measure which is considered by Indian academic regulatory bodies to be reflective of relative importance of a journal within its field, http://en.wikipedia.org/wiki/Impact_factor.

10. UGC: University Grants Commission, the apex academic body of India, http://www.ugc.ac.in/; 10a. University Grants Commission Mandate, http://www.ugc.ac.in/about/mandate.html; 10b. Professional Councils of UGC, AICTE entry: http://www.ugc.ac.in/inside/pcouncil.html#AICTE; 10c. Higher Education in India at a Glance, Feb. 2012, http://www.ugc.ac.in/inside/statistics.html; 10d. AICTE: All India Council of Technical Education: http://www.aicte-india.org/.

11. UGC National Eligibility Test (NET) has only one subject for CS & IT which is Computer Science and Applications: http://www.ugc.ac.in/inside/syllabus.html. The syllabus for Computer Science and Applications subject is available here: http://www.ugc.ac.in/inside/syllabuspdf/87.pdf.

The eligibility tests for Science subjects are conducted by a different organization, Council for Scientific and Industrial Research (CSIR), and that list, http://csirhrdg.res.in/Syllabi_NET.htm, does not have any CS & IT subject(s). But "Common Elementary Computer Science" questions are part of the syllabus of all Science subjects.

12. Pallab De, The State of Engineering in India, techie-buzz.com, May 2011, http://techie-buzz.com/discussions/engineering-colleges-students-india.html.

13. Geeta Anand, India Graduates Millions, but Too Few Are Fit to Hire, The Wall Street Journal, April 2011, http://online.wsj.com/article/SB10001424052748703515504576142092863219826.html.

14. Mark Tarver, Why I am not a Professor, Lambda Associates, 2007, http://www.lambdassociates.org/blog/decline.htm.

15. Computer Science (CS) / Computer Science & Engineering (CSE) and Information Technology (IT) departments in India offer:
- B.E./B.Tech.(CS/CSE), M.E./M.Tech.(CS/CSE)
- B.E./B.Tech.(IT), M.E./M.Tech.(IT)
- B.C.A. and M.C.A. degrees.
  *[B.E. - Bachelor of Engineering, M.E. - Master of Engineering*
  *B.Tech. - Bachelor of Technology, M.Tech. - Master of Technology*
  *B.C.A. - Bachelor in Computer Applications, M.C.A. - Master in Computer Applications.]*

Model syllabus for:
- B.E./B.Tech. in CSE: http://www.aicte-india.org/downloads/mugcomputersc.pdf
- MCA: http://www.aicte-india.org/downloads/mcadegree.pdf

# Appendix A

# Review Remarks of Noted International Academic Publication and Response of Author

After a lot of consideration the author decided to take the unusual step of sharing review remarks for this article by a noted international academic publication focused on CS education. The editor-in-chief of the publication graciously provided permission to share the valuable review remarks of the knowledgeable reviewers in this appendix. The author thanks him and the publication for this kind gesture.

As the author sees it, the stake holders of software education imparted by UGC/AICTE recognised institutions in India are:

1. Students (& Parents): They invest their time and pay the tuition fees.
2. Teachers/Academics: They are supposed to be knowledgeable and do the primary task of imparting appropriate knowledge to students.
3. Employers (Industry): They use the products of the education system (students-turned-graduates) to contribute to economic work and provide a livelihood for these students-turned-graduates.
4. Funding agencies & regulators, namely MHRD (Ministry of Human Resource Development, http://mhrd.gov.in/), UGC & AICTE: They provide the tax payer contributed money for higher education (e.g. as UGC grants) and try to maintain good standards of education. They also look at nationwide issues and society issues like the needs of the country and imparting ethics. Further, they try to promote an environment that will encourage good education (attract good teachers, provide job security to teachers, give students a safe environment, etc.)

The author is of the opinion that this article/paper may not be easily accepted in a forum primarily controlled by one of the stakeholders here, namely the teachers/academics, as it is somewhat critical of them even though the criticism is mainly directed at the system rather than the individual academics. But the article/paper may find a lot of acceptance in forums of some of the other stakeholders especially students, parents, industry and perhaps even MHRD.

This article went through 2 rounds of review with the noted international academic publication. The first round feedback was incorporated in this version of the article, which is what was submitted for the 2nd round review (except for a minor difference in the title of the article). It was not found suitable for the publication in the 2nd round review though the reviewers had some appreciation for the article.

The author views the 2nd round reviewer comments and his response to them as a debate between CS academic viewpoint and industry software developer viewpoint. (The author is not an academic but a software design & development practitioner from the industry who helped out a CS department in an Indian educational institution, a deemed university, to improve the practice of software development in it). He opines that some stakeholders of software education like students, parents and industry practitioners may want to read this debate and that it may contribute to a better understanding of this issue among the public at large. Very importantly, the reviewers provided very knowledgeable comments which contribute significantly to the discourse on the topic. Therefore the author decided to share the review remarks and his response to them in this appendix.

Reviewer: 1
Comments to the Author

> Author: Firstly, thank you very much for your valuable remarks. They have contributed significantly to my understanding of this issue from an international perspective.

Reviewer1: This paper makes an interesting and controversial case for creating career tracks in the Indian CS&IT academia for faculty who, instead of pursuing theoretical research, would opt for establishing academic credentials based on their output of open source software. Much of this seems to reflect from the author's personal journey from being a non-CS graduate, with extensive experience in the CS&IT Software Development industry, and then trying to "fit in" into the academic make-up of an Indian University which seems to be heavily regulated centrally.

> Author: I disagree with a part of the last statement. I consider myself to be an accomplished industry-trained and self-taught software industry technical consultant, who, mainly as an Honorary Faculty/Visiting Faculty, provided free teaching and guidance service to students in software development/engineering and thereby contributed to strengthening the practice of software development/engineering in a CS department in India. I was not and am not interested to "fit in" the academic make-up of any Indian university at all - my intention was to help students learn the practice of software development well, and I believe I succeeded in no small measure in that regard. From my experience of Indian CS academia I had some suggestions to improve the practice of software design & development/engineering in it which I put forward in this article/paper.

Reviewer1: Many issues here: the first and foremost, whether the specific situation in Indian academia is worth publishing in ---publication-name-description-blinded--- with a much broader international audience. On the other hand, given that such situations exist in several countries, this is a good discussion starter to bring a broader awareness to the issues faced and the possible (in my mind, skewed) solution being proposed.

From what I gather, the main problem seems to be in the area of software engineering education.

> Author: Yes, but the whole gamut of software engineering involving design, development, testing etc. and not just a software development process/life-cycle theory course. A published research paper mentioned that in India, "Most universities offer SE as just one of the courses along with other Computer Science courses.", Kirti Garg, Vasudeva Varma, "Software Engineering Education in India: Issues and Challenges," cseet, pp.110-117, 21st Conference on Software Engineering Education and Training, 2008, http://dx.doi.org/10.1109/CSEET.2008.36.

Reviewer1: While, on one hand, most CS&IT faculty in India seem to be primarily devoted to theoretical research, there seems to be a practice among academia of relaxing the qualifications for hiring CS&IT faculty due to the "paucity" of qualified faculty. It is easy to see that the poor quality of Indian CS&IT graduates being prepared for software development jobs in India upon graduation is an artifact of poor quality software engineering education (or a complete lack of it).

> Author: I tend to agree with these views in the Indian context.

Reviewer1: The solution being proposed, to create UGC regulated mandates for separate faculty career tracks, is rather controversial, and perhaps misplaced when one sees it in the larger context of the role of academia. Much of the problem should, and can, be addressed by creating a well qualified pool of faculty in the CS&IT disciplines, software engineering included. Additionally, along with the faculty, to create educational tracks in CS&IT Departments at universities to teach software engineering curricula. Without a presence of these two things, it seems like a proposal to create accommodations for industry professionals to enter non-research tracks in academia, is misplaced and a poor solution.

> Author: It certainly is *not* a proposal focused on creating accommodations for industry professionals to enter non-research tracks in academia. It is a proposal to provide career growth incentive for Indian CS & IT academics to excel in the practice of software development instead of focusing on research publication output and ignoring excellence in practice of software development. Industry professionals being accommodated in a non-research software development career track is a secondary and optional part of what this paper proposes, which in my opinion, has significant value for improving the practice of software development in Indian CS & IT academia.

Reviewer1: This can be detrimental to the intrinsic health and make-up of an entire higher educational system. The case is made, based on faculty in the Performing Arts which tends to be one of the few outliers in this regard. Even in Performing Arts, there is much resistance. Ordinary, day-to-day practitioners of the art seldom attain faculty status even in the Performing Arts.

> Author: In India, it is common to see faculty of the performing arts deliver a performance to the public which leaves students, parents and the public in general in no doubt as to the practical skill of the performing arts faculty. In marked contrast, there is huge amount of doubt in the mind of students, parents and the public in general about the practical software development/engineering skill of most Indian CS & IT academics.

Reviewer1: Besides, there are existing models that accommodate "both" classes of faculty in a single framework that are present outside India that need to be examined. For example, giving academic credit for software artifacts during the promotion and tenure process is widely promoted by the Guidelines published by the Computing Research Association (in the USA). The author should take a look at that.

> Author: It is interesting and it will be good if Indian academic regulations take note of it and provide significant academic credit for software artifacts. I agree very much with the view expressed therein that, "Assessing artifacts requires evaluation from knowledgeable peers." In the Indian context, in my opinion, at least in the short term, it is the software industry which has the capacity to provide enough numbers of knowledgeable peers to evaluate software artifacts produced by Indian CS & IT academia.

Reviewer1: Much of the surveys presented in the paper are about the state (or lack there of) of software engineering education in India. To suggest that a government regulated body create a "practitioner track directed to serving a dimension of a transient and evolving industry" and further put into place specific software-based evaluation metrics for the hiring, promotion, and career advancement of such faculty is a bizarre idea that makes for an excellent blog post, or an opinion piece.

> Author: I humbly submit that as a practitioner of software development I find it bizarre that, in the key regulations that govern Indian CS & IT academia, there is zero career growth incentive to excel in the practice of software development. That, in my opinion, is the key reason for such poor quality of software engineering/software development skills in most Indian CS & IT academics.
>
> Further, I believe that the huge growth of the software design & development field in India is enough reason to seriously consider the suggestion of a specific Indian academic career track which focuses on software design & development. It may be a very strange suggestion for academia in general but the software revolution in the past few decades has changed India and the world quite a bit and Indian academia may need to look at new ways to effectively handle its duties of teaching Indian students the vital skill of software design & development.

Reviewer1: This submission to ---publication-name-blinded---, in this sense, is misplaced.

---

Reviewer: 2
Comments to the Author

> Author: Firstly, thank you very much for your valuable remarks. They have contributed significantly to my understanding of this issue from an international perspective.

Reviewer2: It is good to see a paper addressing issues relating to the relevance and quality of computing education at a national level in India, and the challenges in preparing competent practitioners for the local IT industry. Such a discussion has potential to be of interest to ---publication-name-blinded--- readers. However the solutions proposed fail to take into account several critical issues.
The move beyond computing as CS, EE or IS to a broader set of cognate computing disciplines in a wider family as noted in the ACM 2005 overview report. Therefore institutions need the flexibility to adapt curricula to meet both local conditions and international standards. A highly rigid national framework militates against such adaptability, and thus we see private organisations filling the gaps by providing vocationally focussed certifications. If the core degree learning provides a sound underpinning education, then maybe this is ok?
Shackelford, R., Cassel, L., Cross, J., Davies, G., Impagliazzo, J., Kamali, R., Lawson, E., LeBlanc, R., McGettrick, A., Slona, R., Topi, H. and vanVeen, M. Computing Curricula 2005 The Overview Report including The Guide to Undergraduate Degree Programs in Computing, Joint Task Force ACM, AIS, IEEE-CS, New York, 2005, 46.

> Author: The above document seems to be a very well thought out and well researched document from a North-American perspective. However I do not know how well it can help solve the problem of very poor practice of software design & development in Indian CS & IT academia which is heavily influenced by the regulations and guidelines of UGC & AICTE.

Reviewer2: It needs to be recognised that the nature of the CS/SE divide is historical and long standing, [as is the role of programming in CS] but the divide is arguably one of the strengths of CS and SE that both theory and practice must interrelate in the achieving of outcomes – so education should recognise this in some form. Cf. for instance the discussion below:
Lister, R., Berglund, A., Clear, T., Bergin, J., Garvin-Doxas, K., Hanks, B., Hitchner, L., Reilly, A. L., Sanders, K., Schulte, C. and Whalley, J. Research Perspectives on the Objects-Early Debate. SIGCSE Bulletin, 38, 4 (Dec 2006), 146-165.

> Author: Noted.

Reviewer2: "These distinctions can be traced back to the origins of the discipline, and early schisms "between the logicians and the technicians", (Clark, 2003) depending upon whether one came from a more theoretically oriented mathematical background, or a more practically oriented engineering background".

The role of the SE discipline has always been problematic, and the tensions between the body of knowledge, the evolving nature of practice in the field, what skills should be taught and the nature of the academy and its value systems has always been an issue. But University systems are by their nature international, and local solutions which focus primarily on the vocational teaching mission [with a goal of producing immediately productive 'drones for industry'] without doing equal justice to the research mission, are likely to result in reputational damage to the institution. Cf. the discussion below:
Clear, T. Software Engineering and The Academy: Uncomfortable Bedfellows? SIGCSE Bulletin, 36, 2 (June 2004), 14-15.

> Author: I read some other messages in the above article:
> "Reflecting upon how this experience had enriched his teaching upon his return to the academy, he also noted that few engineering educators possessed any experience of engineering practice."
> ...
> "If we consider medicine as an analogous profession, have not the medical educators themselves completed clinical practice requirements? Would doctors who had never practiced be regarded as credible professors of clinical medicine? Why do we privilege the doctoral qualification over the practice credentials in the case of our software engineering professors?"
...
"Being prepared to recruit ex-practitioners without PhD qualifications, and recognise their value in non-traditional ways may be strategies vital to success in teaching a quality software engineering programme."

In my opinion the above views match the views expressed in my paper.

Reviewer2: But I doubt that measuring an academic's software capability and contribution by open source software production is practicable. (Although for the 'R' statistical software package, newly contributed and specialised statistical modules are formally peer reviewed before acceptance). For instance Open Source is but one mode of software development. What of the skills of developing proprietary software in teams?

> Author: The problem with proprietary software would be availability of source code for reviewers. I guess this would be similar to proprietary research work which is not published in academic research publications and therefore may not contribute to an academic's career growth directly.

Reviewer2: What of domains of application? What of experience of software engineering gained through research involvement with software development firms? Such partnership models are often applied as academics move away from regular software development, and perhaps through supervision of development by students, to a more theoretical, SE process, SE practice or managerial SE research focus.

> Author: The software contribution record that I suggested looked only at software contributions. It can be discussed whether it should be expanded to include some of the above suggestions and the manner in which it should be included.

Reviewer2: The reality of the research teaching divide is also endemic in the academy, as developed in the paper below, and as recently observed by the Business School Accreditation Body AACSB's Blue Ribbon Committee in the report below:

Clear, T., Valuing Computer Science Education Research? [Invited Presentation]. in 6th Baltic Sea Conference on Computing Education Research (Koli Calling 2006), (Koli, Finland, 2006), Uppsala University, Uppsala.
AACSB. Discussion Paper – Relationship Between Research and Teaching, AACSB Blue Ribbon Committee on Accreditation Quality 2011, 1-5.

> Author: Noted.

Reviewer2: In other systems when hiring academics, their skill sets are carefully considered against the needs of the department. It is unlikely that an academic with no knowledge of software engineering practice would be hired to teach a software engineering course, although they may teach a math course or a course in theoretical CS.

Author: Unfortunately what Clear,T. wrote (mentioned earlier), "few engineering educators possessed any experience of engineering practice" applies very well to Indian CS & IT academia. Career growth incentive to them to improve in software engineering practice may lead them to make efforts to do so. Otherwise it is natural that they will be attracted only towards producing research publications as that provides career growth.

Reviewer2: As presented, while there is active debate in many of these areas, the paper does not fully address the issues [disciplinary, institutional, cultural, political] that are obstacles to implementing such a nationwide programme of change in the Indian Higher Education System. It poses a challenging set of questions in the context of the author and his perspective on computing education in his country, and how practice and theory gap might be reduced. To that extent the issues are important to all computing educators, so a debate could certainly be had. As it stands unfortunately the paper lacks the dimensions to productively lead that debate.

With a more considered set of proposals and a wider understanding of the academic setting and how to effect change in a complex context with local and global dimensions, there could be a stronger candidate for publication.